\documentclass[sigconf,screen]{acmart}

\usepackage{todonotes}
\usepackage{siunitx}
\usepackage{balance}

\AtBeginDocument{%
  }

\copyrightyear{2024}
\acmYear{2024}
\setcopyright{acmlicensed}\acmConference[UbiComp Companion '24]{Companion of the 2024 ACM International Joint Conference on Pervasive and Ubiquitous Computing }{October 5--9, 2024}{Melbourne, VIC, Australia}
\acmBooktitle{Companion of the 2024 ACM International Joint Conference on Pervasive and Ubiquitous Computing (UbiComp Companion '24), October 5--9, 2024, Melbourne, VIC, Australia}
\acmDOI{10.1145/3675094.3678480}
\acmISBN{979-8-4007-1058-2/24/10}

\begin{document}

\title{OpenEarable ExG: Open-Source Hardware for Ear-Based Biopotential Sensing Applications}

\author{Philipp Lepold}
\email{lepold@teco.edu}
\affiliation{%
  \institution{Karlsruhe Institute of Technology}
  \city{Karlsruhe}
  \country{Germany}
}

\author{Tobias Röddiger}
\email{tobias.roeddiger@kit.edu}
\affiliation{%
  \institution{Karlsruhe Institute of Technology}
  \city{Karlsruhe}
  \country{Germany}
}

\author{Tobias King}
\email{tobias.king@kit.edu}
\affiliation{%
  \institution{Karlsruhe Institute of Technology}
  \city{Karlsruhe}
  \country{Germany}
}

\author{Kai Kunze}
\email{kai@kmd.keio.ac.jp}
\affiliation{%
  \institution{Keio Media Design}
  \city{Keio}
  \country{Japan}
}

\author{Christoph Maurer}
\email{christoph.maurer@uniklinik-freiburg.de}
\affiliation{%
  \institution{University of Freiburg}
  \city{Freiburg}
  \country{Germany}
}

\author{Michael Beigl}
\email{michael.beigl@kit.edu}
\affiliation{%
  \institution{Karlsruhe Institute of Technology}
  \city{Karlsruhe}
  \country{Germany}
}

\renewcommand{\shortauthors}{Philipp Lepold et al.}

\begin{abstract}
While traditional earphones primarily offer private audio spaces, so-called ``earables'' emerged to offer a variety of sensing capabilities.
Pioneering platforms like OpenEarable have introduced novel sensing platforms targeted at the ears, incorporating various sensors. The proximity of the ears to the eyes, brain, and facial muscles has also sparked investigation into sensing biopotentials.
However, currently there is no platform available that is targeted at the ears to sense biopotentials.
To address this gap, we introduce \textit{OpenEarable ExG} - an open-source hardware platform designed to measure biopotentials in and around the ears. OpenEarable ExG can be freely configured and has up to 7 sensing channels.
We initially validate OpenEarable ExG in a study with a left-right in-ear dual-electrode montage setup with 3 participants.
Our results demonstrate the successful detection of smooth pursuit eye movements via Electrooculography (EOG), alpha brain activity via Electroencephalography (EEG), and jaw clenching via Electromyography (EMG). OpenEarable ExG is part of the OpenEarable initiative and is fully open-source under MIT license.
\end{abstract}

\ccsdesc[500]{Hardware~Emerging technologies}
\ccsdesc[500]{Hardware}
\ccsdesc[500]{Human-centered computing~Ubiquitous and mobile computing systems and tools}
\ccsdesc[500]{Human-centered computing~Ubiquitous and mobile devices}

\keywords{open-source; earables; hearables; bio-potential; Electrooculography; Electroencephalography; Electromyography; EOG; EEG; EMG}

\maketitle

\begin{figure*}[!t]
    \centering
    \includegraphics[width=\linewidth]{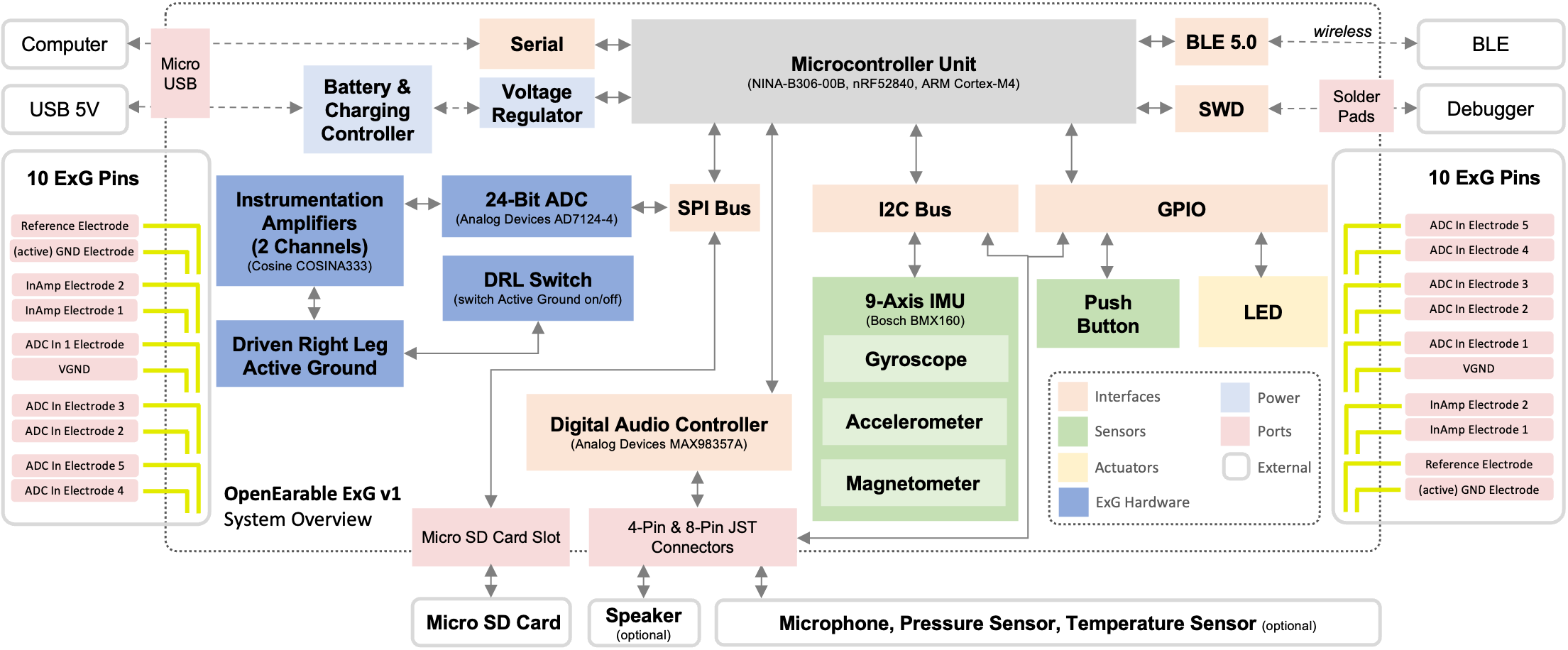}
    \caption{An overview of the system architecture of OpenEarable ExG which builds upon OpenEarable v1.3 \cite{roddiger2022openearable}.}
    \label{fig:figure1}
\end{figure*}

\section{Introduction and Related Work}
The head offers unique opportunities for integrating sensing and actuation~\cite{amft2015making}. As the primary location for many of our senses - including sight, sound, touch, taste, and smell - the head is an intriguing focal point for innovative wearables, such as Earables. Earables are earphones that offer functionalities beyond audio in- and output \cite{roddiger2022sensing}.
\citet{roddiger2022sensing} have shown that in past earables research, various sensing principles have been investigated including motion, audio, optical, and environmental sensor.
Another sensing principle that has been investigated broadly on the ears are biopotentials including Electronencephalography (EEG) \cite{kidmose2013study}, Electrooculography (EOG) \cite{hladek2018real,manabe2013conductive}, Electrocardiography (ECG) \cite{von2017hearables}, and Electromyography (EMG) \cite{bi2017toward}.
Phenomena that can be detected using biopotentials measured at the ears include EEG standard protocols such as SSVEP~\cite{ahn2018wearable} or AAR~\cite{bleichner2017concealed}. Concrete applications based on ear biopotentials include seizures detection \cite{gu2017comparison}, jaw clenching monitoring \cite{dong2016new}, and tracking eating \cite{bi2017toward}.

While there are research platforms such as OpenEarable \cite{roddiger2022openearable}, eSense \cite{kawsar2018earables}, and the OpenBCI-based Open ExG Headphones \cite{knierim2023openbci} for research on different ear-based sensing applications, there is currently no open-source research platform targeted at the ears to measure biopotentials in-ear.
Therefore, in this paper we introduce \textit{OpenEarable ExG}, an open-source platform to measure biopotentials. 
Our hardware is based on OpenEarable v1.3 \cite{roddiger2022openearable} and OpenBCI \cite{openbci_cyton}. 
We provide the schematics, PCB design files and CAD models for a device that sits on the neck able to record biopotentials.
We support up to 7 sensing channels (incl. 1 reference) plus ground. 
In a study, we showcase our system for EEG, EMG and EOG applications. The device can also be extended freely via two connectors and 2$\times$10 jumper pins.
Overall, the contributions of our paper are:
\begin{enumerate}
    \item \textit{OpenEarable ExG:} an open-source, extensible platform for sensing biopotentials in and around the ears
    \item an evaluation with 3 participants showing that OpenEarable ExG can measure EEG, EMG, and EOG phenomena
\end{enumerate}


\section{Design Principles}
Our vision is to establish the \textit{OpenEarable ExG} as a versatile open-source platform. To achieve this, we adhere to the best practices of the Open-Source Hardware Association (OSHWA) \cite{oshwa} by (i) sharing all original hardware design source files and ready-to-view outputs (e.g. PDFs); (ii) using freely available tools, common production processes and widely available components for all designs; (iii) open-sourcing all firm- and software; (iv) providing photos and instructions how to make and use the hardware; (v) licensing all resources under MIT license\footnote{\url{https://github.com/OpenEarable/open-earable-ExG}}.
Components are selected based on proven parts from related hardware which are known to have good availability, primarily based on OpenEarable 1.3 \cite{roddiger2022openearable}.


\section{Hard- and Software}
The following sections describe the hard- and software, and the mechanical assembly of OpenEarable ExG as well as the extensibility options of the platform via different connectors.

\begin{figure*}[!t]
    \centering
    \includegraphics[width=\linewidth]{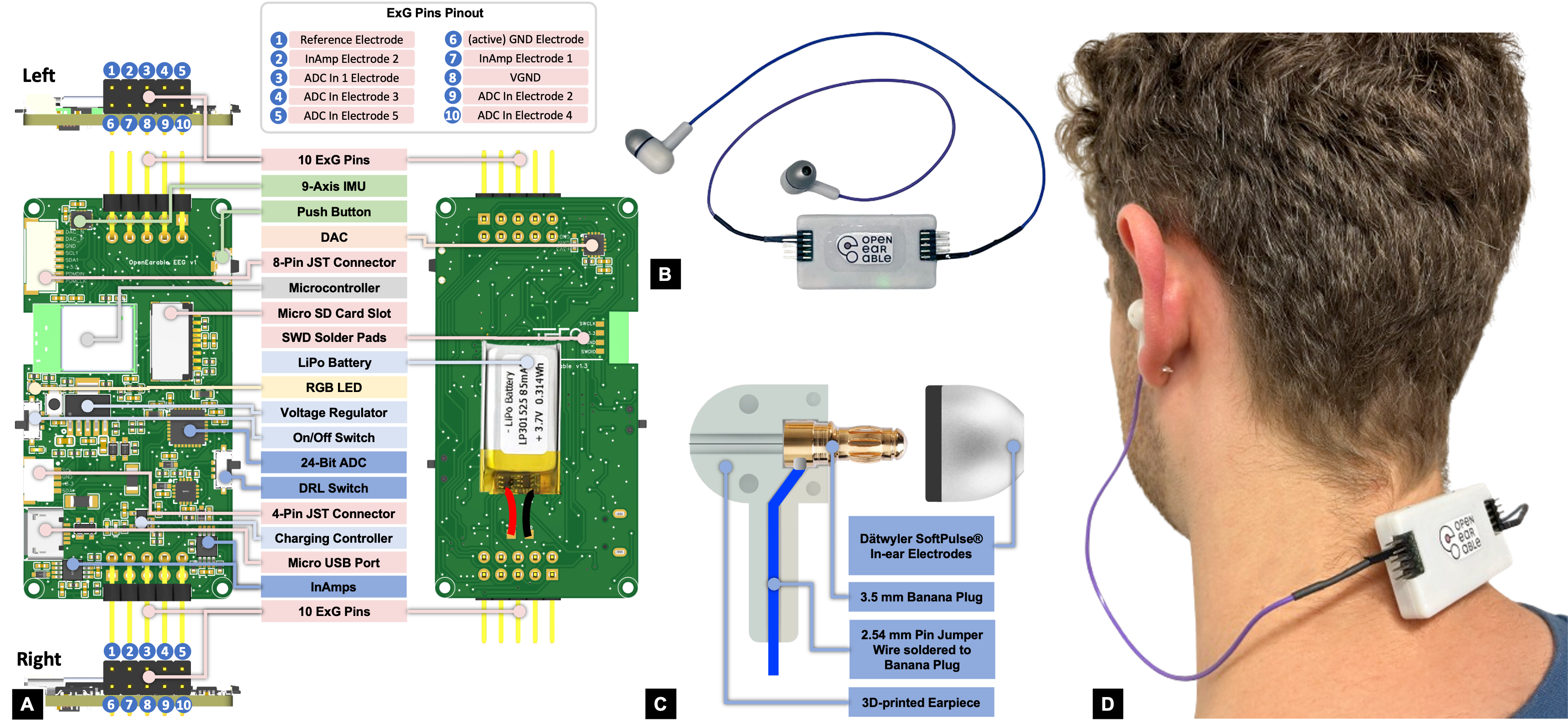}
    \caption{(A) PCB and components of OpenEarable ExG; (B) 3D-printed in-ear plugs based on Dätwyler SoftPulse in-ear electrodes \cite{datwyler_softpulse_products} (C) assembled device with 3D-printed housing and custom in-ear electrodes; (D) user wearing OpenEarable ExG.}
    \label{fig:figure2}
\end{figure*}

\subsection{OpenEarable Core}
At the core of OpenEarable ExG is the ``OpenEarable 1.3'' by \citet{roddiger2022openearable}. It is based on the NINA-B306-00B module containing the Nordic nRF52840 \cite{ublox_nina_b3}, a single core ARM Cortex-M4 microcontroller. It also integrates the Bosch BMX160 9-axis IMU (accelerometer, gyroscope, magnetometer)  \cite{bosch_bmx160} able to filter out motion artifacts, a charging controller (Microchip Technology MCP73831T \cite{microchip_mcp73831}), a slide switch to turn the device on and off, an RGB status LED, and a charging indicator LED. Lastly, OpenEarable ExG also has a digital-to-analog converter (MAX98357AETE \cite{analog_devices_max98357}) to output audio signals and a microSD card slot to record data at high sampling rates locally.

\subsection{ExG Hardware}
The following subsections introduce the ExG specific hardware of OpenEarable ExG.

\subsubsection{Analog-to-Digital Converter}
The most critical component of OpenEarable ExG is the analog digital converter.
We chose the Analog Devices AD7124-4, 4-channel (4 differential/7 pseudo differential inputs), 24-bit sigma-delta ADC \cite{analog_devices_ad7124_4} mainly due to its high resolution (similar to state-of-the-art devices like the OpenBCI \cite{openbci_cyton}), its very high gain stage, and its built-in 50 Hz/60 Hz rejection of up to 130 dB. This enables OpenEarable ExG to directly use the sampled data on device, without any latency due to digital filtering, for example for on-device classification of different ExG phenomena. 
The ADC needs to have sufficiently good resolution to sample the small voltage differences and, together with an in-amp gain of \textasciitilde 50, in a 3.3 V system at least 17 bit are necessary so that artifacts are not suppressed by quantization \cite{teplan_fundamentals_2002}.
Among all alternatives, the AD7124-4 is (as of today) also consistently in stock and relatively cheap (approx. USD 12.73).
The AD7124-4 features three power modes, including one low power mode which uses less than 1 mW, making it an excellent choice for a wearables.
ADC AIN0 channel is connected to the virtual ground of 1.65 V (VDD/2), allowing it to measure the difference between the reference pin of OpenEarable ExG and Electrode 1 \& 2 (ADC channels AIN1 and AIN2) which both are amplified by instrumentation amplifiers before reaching the ADC. This leaves ADC channels AIN3 - AIN7 unoccupied and accessible via the ExG pin headers (see \autoref{fig:figure2}.A for the pinout).
\subsubsection{Instrumentation and Operational Amplifier}
To achieve a good common mode rejection (120 dB, V\textsubscript{CM} < 5\si{\volt}) between the electrodes and even better 50 Hz/60 Hz rejection, two Cosine COSINA333 \cite{cosina333} rail-to-rail instrumentation amplifiers are incorporated in OpenEarable ExG, between Electrode 1 \& 2 and the Analog-to-Digital-Converter.
These in-amps have been chosen due to their well suited performance and capability to operate with 3.3 V as well as their ability to set the gain using only one resistor, minimising space on the PCB. The gain of \textasciitilde 50 is set through a 2.2 k\si{\ohm} resistor. This gain alongside with the 24-bit resolution in a 3.3 V powered device allows for a theoretical minimal detectable voltage difference as low as \textasciitilde4 nV between the reference and measurement electrodes.

To provide a stable virtual ground of 1.65 V (VDD/2), which enables measuring positive as well as negative differences between the reference and measurement electrode, a voltage follower circuit is incorporated, using the Willsemi WS72324Q operational amplifier. The WS72324Q has rail-to-rail capabilities, operates at 3.3 V and is available in a tiny four-opamp-in-one QFN3x3-16L package, making it ideal for OpenEarable ExG.
The Driven-Right-Leg active ground also needs two opamp circuits, leaving one circuit unused.

\subsubsection{Driven-Right-Leg Active Ground}
To increase the 50 Hz/60 Hz rejection even further and provide better signal quality, a Driven-Right-Leg (DRL) circuit is integrated into OpenEarable ExG and available via the (active) GND pin, as seen in \autoref{fig:figure2}.A. This circuit senses the 50 Hz/60 Hz noise at the instrumentation amplifier stage and inverts this signal to feed it back to the body via the GND electrode reducing/canceling the 50 Hz/60 Hz noise \cite{acharya2011improving}. This is, simplified, comparable to noise cancellation in modern headphones. The DRL functionality can be disabled via the DRL switch (refer to \autoref{fig:figure2}.A) so that the GND pin only outputs the virtual ground voltage of 1.65 V (VDD/2). 

\subsubsection{Pin-Assignment}
OpenEarable ExG features 10 header 2.54 mm pitch jumper pins on each side. The header pins provide easy access to connect electrodes. 
The pinout is detailed in \autoref{fig:figure2}.A.

\subsubsection{Earpiece and Mechanical Design}
The earpieces of OpenEarable ExG are custom made. We designed a 3D-printed enclosure into which a male 3.5 mm banana plug with cable is glued into (see \autoref{fig:figure2}.C). The eartips are Dätwyler SoftPulse in-ear electrodes \cite{datwyler_softpulse_products} which are simply plugged onto the connector. This design is lightweight, comfortable to wear, looks like standard earphones and the electrodes can be easily swapped for different users.

The enclosure of OpenEarable ExG (see \autoref{fig:figure2}.B) is 3D-printed. It leaves all necessary ports and IO available from the outside and encloses the LiPo battery.
All parts are printed with a Formlabs 3+ SLA 3D printer and the CAD files are open-source.

\subsubsection{Extensibility} OpenEarable ExG features 10 symmetrical header pins on each side. These provide easy access via jumper wires to the two inputs that have an instrumentation amplifier stage and their reference, the five other inputs that are directly attached to the ADC as well as the virtual ground and a driven right leg output.
Moreover, we ensure that the platform remains extensible by adding the 8-pin and 4-pin connector of OpenEarable 1.3 \cite{roddiger2022openearable} which supports PDM microphones, I2C devices, analog audio output, and general-purpose IO. This will enable researcher to integrate OpenEarable ExG in diverse research settings, e.g. including audio applications or the integration of a variety of biosensing principles.

\begin{figure*}[!t]
    \centering
    \includegraphics[width=\linewidth]{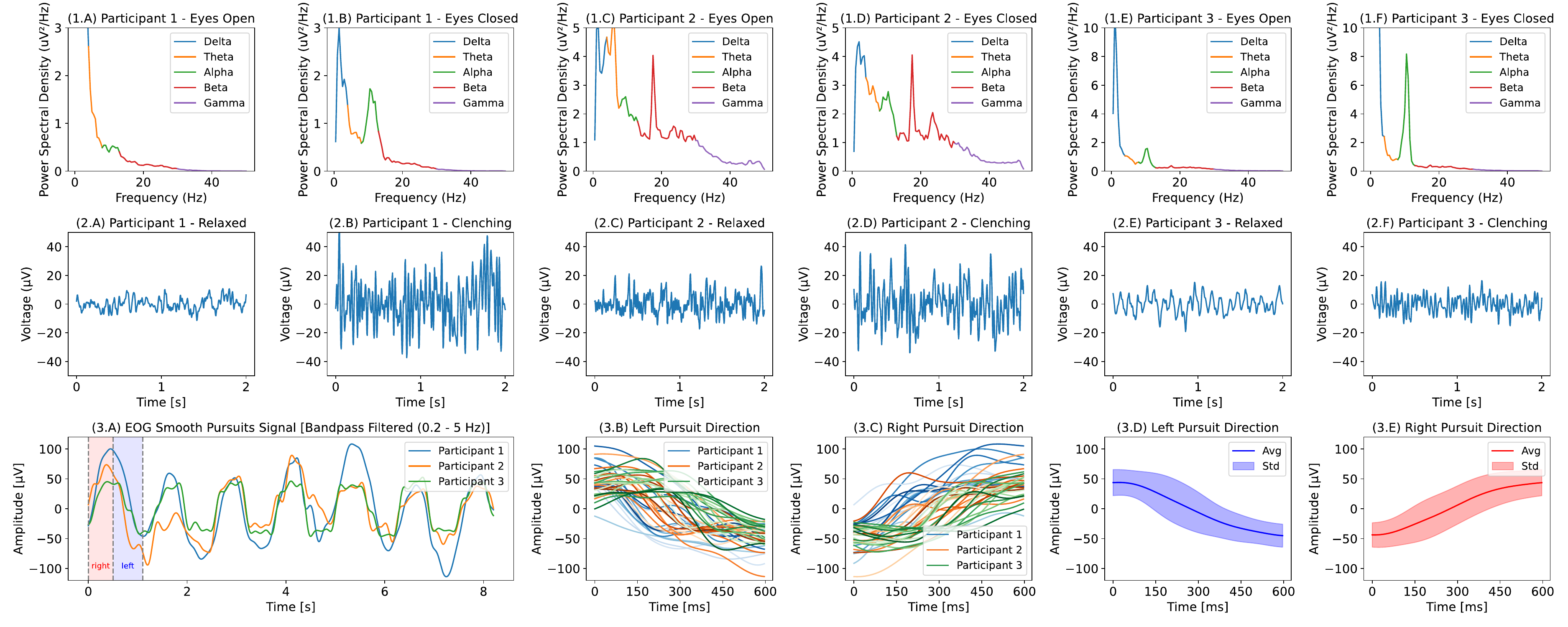}
    \caption{(1.A) - (1.F) display brain activity measured via ear EEG for three participants, with eyes open and closed, revealing increased alpha activity when eyes are closed. (2.A) - (2.F) illustrate the rise in signal amplitude measured by ear EMG during jaw muscle clenching for the same participants. (3.A) - (3.E) present smooth pursuit signals measured via ear EOG, showing signal deflection in opposite directions based on the participants' gaze direction.}
    \label{fig:figure3}
\end{figure*}
\subsection{Software}
OpenEarable ExG extends the firmware of OpenEarable 1.3 which is based on Arduino \cite{roddiger2022openearable}. For data recording, different Python scripts are available to record data via Bluetooth or wired via USB Serial. Over serial, sampling rates of up to 19,200 SPS are possible (limited by ADC hardware), over BLE sampling rates of up to 500 SPS are possible. The firmware and recording software are available in the GitHub repository of OpenEarable ExG\footnotemark[\value{footnote}].

\section{Applications}
To demonstrate the versatility of OpenEarable ExG, we present its effectiveness in three applications: brain activity sensing via EEG, jaw clenching detection via EMG and eye tracking via EOG. We recruited three male participants through a sample of convenience (mean age 24.0 years, SD = 1.4 years). The in-ear electrodes were placed in the left and right ears, with the right ear serving as reference and the left ear as measurement electrode (using Electrode 1 with in-amp stage). No ground electrode was used. The results of this study can be seen in \autoref{fig:figure3}. 

\subsection{EEG: Brain Activity Sensing}
Previous studies have established that alpha band power in brain activity increases when the eyes are closed \cite{hohaia2022occipital}. Accordingly, we asked participants to fixate on a point for two minutes with their eyes open on a computer screen, and then close their eyes for two minutes. \autoref{fig:figure3}.1 illustrates a clear increase in alpha power for all three participants when comparing the eyes-open condition to the eyes-closed condition. This suggests OpenEarable ExG offers the foundation for the exploration of wearable EEG scenarios.

\subsection{EMG: Jaw Clenching}
The ears are situated near several muscular structures of the head and face. Jaw clenching, which often occurs during sleep or high stress, involves excessive pressing of the teeth. In our study, participants were asked to clench their teeth for 10 seconds as hard as they comfortably can. \autoref{fig:figure3}.2 presents sample plots from each participant, highlighting the noise produced by the flexing of the masseter muscle, which is located close to the in-ear electrodes. The the differences compared to the resting state signal are obvious across all participants.
\
\subsection{EOG: Eye Tracking}
Eye gaze and in particular EOG is an interesting parameter to diagnose imbalance disorders \cite{newman2021detecting}.
To understand gaze tracking capabilities of OpenEarable ExG, participants were seated with their heads positioned in a chin rest, 40 cm away from a 23-inch 1920 $\times$ 1080 px screen, which was centered at eye level. By following a point on the screen moving horizontally, participants performed a smooth pursuit with an angle of around 52.8° in 0.5 s both to the left and right respectively, repeating this 15 times per side. 
The resulting signal deflections are shown in \autoref{fig:figure3}.3. 
It can be seen that OpenEarable ExG is perfectly capable of measuring EOG signals across all study participants.

\section{Conclusion}
In this paper, we introduced OpenEarable ExG, an open-source hardware platform for ear-based biopotential sensing applications. The key contributions of this work include:
Our results show that OpenEarable ExG is capable of detecting standard EEG phenomena like alpha brain activity, jaw clenching via EMG, and smooth pursuit eye movements via EOG, all using a simple in-ear electrode configuration. This versatility highlights the power and ease of use of ear-based biopotential sensing using OpenEarable ExG for a wide range of applications, including health monitoring, human-computer interaction, and cognitive state assessment.
By making OpenEarable ExG fully open-source under the MIT license, we aim to lower the barrier to entry for researchers and developers interested in exploring ear-based biosensing. The extensible nature of the platform allows for easy customization and integration with other sensors or systems, opening up possibilities for novel multi-modal sensing applications.

\begin{acks}
This work was partially funded by the Deutsche Forschungsgemeinschaft (DFG, German Research Foundation) – GRK2739/1 - Project Nr. 447089431 – Research Training Group: KD2School – Designing Adaptive Systems for Economic
Decisions and by the Carl-Zeiss-Stiftung (Carl-Zeiss-Foundation) as part of the project "JuBot - Staying young with robots".
\end{acks}

\bibliographystyle{ACM-Reference-Format}
\balance
\bibliography{bibliography}


\end{document}